\documentclass[twocolumn,showpacs,aps,prl,superscriptaddress]{revtex4}

\usepackage{graphicx}
\usepackage{dcolumn}
\usepackage{amsmath}
\usepackage{epsfig}

\begin{document}

\title{
  {\large 
    \bf \boldmath 
    $\Delta \Gamma_d$: A Forgotten Null Test of the Standard Model
  }
}

\author{Tim Gershon}
\affiliation{Department of Physics, University of Warwick, Coventry CV4 7AL, United Kingdom}

\date{\today}

\begin{abstract}
The recent measurement of an anomalous like-sign dimuon asymmetry by the D0
collaboration has prompted theoretical speculation on possible sources of
physics beyond the Standard Model that may affect lifetimes and lifetime
differences in neutral $B$ meson systems.
One observable that deserves closer attention is the width difference in
the $B^0_d$ system, $\Delta \Gamma_d$.
Since the Standard Model prediction for this quantity is well below $1\,\%$, 
it serves as a ``null test'' whereby the measurement of a larger value would
cleanly reveal the presence of new physics. 
Methods to measure $\Delta \Gamma_d$ at current and future experiments are
reviewed and an attractive new approach is proposed.
\end{abstract}


\maketitle

Recently, the D0 collaboration have measured an anomalous like-sign dimuon
asymmetry~\cite{Abazov:2010hv,Abazov:2010hj}, which could originate from $CP$
violation effects in either or both of the $B^0_d$--$\bar{B}^0_d$ and
$B^0_s$--$\bar{B}^0_s$ systems.
Since the Standard Model (SM) predictions for $CP$ violation in mixing in both
systems are ${\cal O}(10^{-3})$ or smaller~\cite{Lenz:2006hd}, a significant
effect at the percent level would be a clear signal of non-SM contributions.
When combined with other constraints on $CP$ violation in
$B^0_d$--$\bar{B}^0_d$~\cite{Aubert:2006nf,Nakano:2005jb} and
$B^0_s$--$\bar{B}^0_s$~\cite{Abazov:2007nw} mixing, it appears most likely
that the effect is not solely due to either neutral meson system alone, but
rather that anomalous $CP$ violation effects may be manifest in
both~\cite{Abazov:2010hv,Abazov:2010hj,Ligeti:2010ia}.
Despite the many experimental studies of the $B_d^0$ system over the past
decade, the possibility of such effects is not ruled out (see for example
Ref.~\cite{Antonelli:2009ws} and references therein).

Although improved experimental measurements are certainly necessary, the D0
result has prompted a great deal of theoretical speculation on sources of
physics beyond the Standard Model (``new physics'') that could be responsible
for the effect.
One interesting possibility is that new physics may cause the neutral $B$
meson lifetimes and lifetime differences to differ from their SM
values~\cite{Dighe:2007gt,Dighe:2010nj,Bauer:2010dg,Bai:2010kf}.
Such scenarios -- in particular the possibility that the width differences
$\Delta \Gamma_q$ ($q=d,s$) between the eigenstates of the effective weak
Hamiltonian in the $B^0_q$--$\bar{B}^0_q$ systems may be enhanced above their
SM values -- have not previously been widely considered.
In particular, with little exception~\cite{Dighe:2001gc,Anikeev:2001rk}, 
the utility of $\Delta \Gamma_d$ to test the SM has been largely overlooked. 

To test new physics models that affect $\Delta \Gamma_q$ it is necessary to
have improved measurements of the parameters of the neutral $B$ meson systems,
and to have precise SM predictions to compare against.  
A useful class of tests is those referred to as ``null tests'', in which the
Standard Model prediction is vanishingly small compared to the experimental
sensitivity.
The search for $CP$ violation in neutral $B$ meson mixing itself falls into
this class of tests.
Among the $CP$-conserving mixing parameters, however, there is also one
parameter which is very small in the SM, namely $\Delta \Gamma_d$,
for which the prediction is~\cite{Lenz:2006hd}
\begin{eqnarray}
  \frac{\Delta \Gamma_d^{\rm SM}}{\Delta M_d^{\rm SM}} & = &
  \left( 52.6 \,^{+11.5}_{-12.8} \right) \times 10^{-4} \, , \\
  \Delta \Gamma_d^{\rm SM} & = & 
  \left( 26.7 \,^{+5.8}_{-6.5} \right) \times 10^{-4} \ {\rm ps}^{-1} \, , \\
  \frac{\Delta \Gamma_d^{\rm SM}}{\Gamma_d^{\rm SM}} & = & 
  \left( 40.9 \,^{+8.9}_{-9.9} \right) \times 10^{-4} \, ,
\end{eqnarray}
where the second and third results are obtained using the experimental values
of the mass difference and average lifetime in the $B^0_d$--$\bar{B}^0_d$
system, 
$\Delta M_d^{\rm exp} = (0.507 \pm 0.004) \ { \rm ps}^{-1}$ and 
$1/\Gamma_d^{\rm exp} = \tau(B_d^0)^{\rm exp} = (1.530 \pm 0.009) \ {\rm ps}$~\cite{Amsler:2008zz}.
This involves an assumption that there are no new physics contributions
to $\Delta M_d^{\rm exp}$ and $\Gamma_d^{\rm exp}$, 
{\it i.e.} that $\Delta M_d^{\rm exp} = \Delta M_d^{\rm SM}$ and
$\Gamma_d^{\rm exp} = \Gamma_d^{\rm SM}$.
The former is indeed confirmed to good precision by global fits to the
Cabibbo-Kobayashi-Maskawa (CKM)~\cite{Cabibbo:1963yz,Kobayashi:1973fv}
Unitarity Triangle (see, for example, Refs.~\cite{Charles:2004jd}
and~\cite{Bona:2006ah}).
In any case, plausible new physics effects in the values of $\Delta M_d$ and
$\Gamma_d$ cannot change the conclusion that 
$\Delta \Gamma_d^{\rm SM}/\Gamma_d^{\rm SM} \ll {\cal O}(10^{-2})$.

It is striking that there are relatively few measurements of $\Delta \Gamma_d$
listed by the Particle Data Group~\cite{Amsler:2008zz} and
the Heavy Flavour Averaging Group~\cite{TheHeavyFlavorAveragingGroup:2010qj}.
The world average, based mainly on a single measurement from
BaBar~\cite{Aubert:2003hd,Aubert:2004xga} is
${\rm sign}({\rm Re}\,\lambda_{CP})\Delta \Gamma_d/\Gamma_d = 0.009 \pm 0.037$.
BaBar use a notation in which ${\rm sign}({\rm Re}\,\lambda_{CP})$ is expected
to be $+1$ in the SM (where it is, to a good approximation, equal to 
${\rm sign}(\cos2\beta)$, where $\beta$ is one of the CKM Unitarity Triangle 
angles).  
As discussed below, $\Delta \Gamma_d$ may be quite challenging to measure, but
nevertheless its determination should be possible for the existing
$B$-factory experiments, BaBar~\cite{Aubert:2001tu} and Belle~\cite{:2000cg},
the Tevatron experiments CDF~\cite{Abe:1988me} and D0~\cite{Abazov:2005pn},
the CERN LHCb experiment~\cite{Alves:2008zz} and any future Super Flavour
Factory~\cite{Adachi:2008da,Bona:2007qt,Browder:2008em}. 
Even discounting any potential sensitivity to new physics, $\Delta \Gamma_d$
is a fundamental parameter of the neutral $B$ meson system and should be
measured as precisely as possible.

As a brief digression, it is interesting to consider the situation in the
$D^0$--$\bar{D}^0$ system (see Ref.~\cite{Artuso:2008vf} for a detailed 
review). 
Before the first evidence of charm oscillations was discovered in
2007~\cite{Aubert:2007wf,Staric:2007dt}, 
the SM values of the mixing parameters 
$x_D = \Delta m_D / \Gamma_D$ and $y_D = \Delta \Gamma_D / 2\Gamma_D$ were
generally believed to be ${\cal O}(10^{-3})$.
However, theoretical re-evaluations have shown that values as large as the
experimental measurements, which are ${\cal O}(10^{-2})$, cannot be ruled out
within the Standard Model.
To resolve the situation, more precise measurements and improved theoretical
calculations are required.
The theoretical situation for the neutral $B$ mesons is, however, somewhat
better compared to that for charm mesons, due to the different regime with
regard to QCD effects.

In order to examine how $\Delta \Gamma_d$ may be determined experimentally,
consider the time-dependent decay rates of a neutral $B$ meson that is
initially (at time $\Delta t = 0$) tagged as $\bar{B}^0_q$ or $B^0_q$ to a
final state $f$.
Assuming $CPT$ conservation, and neglecting corrections arising from $CP$
violation in mixing (discussed further below), these are given
by~\cite{Dunietz:2000cr}
\begin{widetext}
\begin{eqnarray}
  \label{B0barRate}
  \Gamma_{\bar{B}_q^0 \to f} (\Delta t) & = &
  {\cal N}_f 
  \frac{e^{-| \Delta t | / \tau(B_q^0)}}{4\tau(B_q^0)}
  \left[ 
    \cosh\left(\frac{\Delta \Gamma_q \Delta t}{2}\right) +
    S_f \sin(\Delta m_q \Delta t) -
    C_f \cos(\Delta m_q \Delta t) +
    {\cal A}^{\Delta\Gamma}_f \sinh\left(\frac{\Delta \Gamma_q \Delta t}{2}\right)
  \right] \, , \\
  \Gamma_{B_q^0 \to f} (\Delta t) & = &
  \label{B0Rate}
  {\cal N}_f 
  \frac{e^{-| \Delta t | / \tau(B_q^0)}}{4\tau(B_q^0)}
  \left[
    \cosh\left(\frac{\Delta \Gamma_q \Delta t}{2}\right) -
    S_f \sin(\Delta m_q \Delta t) +
    C_f \cos(\Delta m_q \Delta t) +
    {\cal A}^{\Delta\Gamma}_f \sinh\left(\frac{\Delta \Gamma_q \Delta t}{2}\right)
    \right] \, ,
\end{eqnarray}
\end{widetext}
where 
\begin{eqnarray}
  S_f = \frac{2\, {\rm Im}(\lambda_f)}{1 + |\lambda_f|^2} \, , 
  & & 
  C_f = \frac{1 - |\lambda_f|^2}{1 + |\lambda_f|^2} \, , \\
  {\cal A}^{\Delta\Gamma}_f = -\frac{2\, {\rm Re}(\lambda_f)}{1 + |\lambda_f|^2} \, , 
  & {\rm and} & 
  \lambda_f = \frac{q}{p}\frac{\bar{A}_f}{A_f} \, .
\end{eqnarray}
Note that $\left( S_f \right)^2 + \left( C_f \right)^2 + 
\left( {\cal A}^{\Delta\Gamma}_f \right)^2 = 1$ by definition.
The parameters $\bar{A}_f$ and $A_f$ are the complex amplitudes for the decay
of a $\bar{B}^0_q$ and a $B^0_q$ to the final state $f$, respectively.
The parameters $q$ and $p$ describe the eigenstates of the effective weak
Hamiltonian of the $B^0_q$--$\bar{B}^0_q$ system
(in the absence of $CP$ violation in mixing, $\left|q/p\right| = 1$).
The constant ${\cal N}_f$ is a normalisation factor.
The equations above are identical for $B$ meson production in any environment,
but in coherent production of $B\bar{B}$ pairs at $e^+e^-$ $B$ factory
colliders the quantum entanglement allows flavour tagging at any time so that
the range of $\Delta t$ is $-\infty < \Delta t < \infty$, whereas $B$ mesons
produced in hadron collisions (or at higher energy $e^+e^-$ colliders) are
tagged at their production point ($0 < \Delta t < \infty$). 
Note that in contrast to the $CP$-violating asymmetry parameters $S_f$ and
$C_f$,  the parameter ${\cal A}^{\Delta\Gamma}_f$ appears with the same sign
in both $\bar{B}^0_q$ and $B^0_q$ decay time distributions.

Focussing now on the $B^0_d$ system, and assuming that 
$y_d = \Delta \Gamma_d / 2\Gamma_d \ll 1$, the following substitutions can be
made 
\begin{eqnarray}
  \cosh\left(\frac{\Delta \Gamma_d \Delta t}{2}\right) & \approx & 1 \, , \\
  \sinh\left(\frac{\Delta \Gamma_d \Delta t}{2}\right) & \approx & 
  y_d \, \Gamma_d \, \Delta t \, , \label{sinhApprox}
\end{eqnarray}
where terms of ${\cal O}(y_d^2)$ have been neglected.
The best channels to measure $\Delta \Gamma_d$ with are therefore those with
large branching fractions and large, well-known values of 
${\cal A}^{\Delta\Gamma}_f$. 
Experimentally, it is also desirable to use final states that are advantageous
from the point of view of minimising systematic errors -- specifically, those
with relatively clean signals and with accessible control samples with
similar topologies.
These considerations suggest that decays mediated by the $b \to c\bar{c}s$
quark-level transition and involving $J/\psi$ mesons in the final state will
be most suitable.

Since the term involving ${\cal A}^{\Delta\Gamma}_f$ is the same for
$\bar{B}^0_d$ and $B^0_d$ decays, $y_d$ can be determined in untagged analyses
({\it i.e.} without determining the initial flavour of the decaying $B$ meson).
With the approximations above, the untagged decay rate is
\begin{equation}
  \label{untaggedRate}
  \begin{array}{lcr}
    \multicolumn{2}{l}{
      \Gamma_{\bar{B}^0_d \to f}(\Delta t) + \Gamma_{B^0_d \to f}(\Delta t)
      \propto
    } & \hspace{20mm} \\
    \phantom{\hspace{20mm}} & \multicolumn{2}{r}{
      e^{-| \Delta t | / \tau(B_d^0)}
      \left( 
        1 + {\cal A}^{\Delta\Gamma}_f y_d \, \Gamma_d \, \Delta t
      \right) \, .
    }
  \end{array}
\end{equation}
Noting that $e^{X(1+\epsilon)} \approx e^X ( 1 + X\epsilon )$, this result
shows that (i) the effect of $y_d \neq 0$ is that the effective lifetimes
measured in decays to different final states (with different values of 
${\cal A}^{\Delta\Gamma}_f$) can differ; (ii) any attempt to measure $y_d$
from a single final state will suffer from large systematic uncertainties.

Bearing this in mind, there are four main possible approaches that can be used
to determine $y_d$.
\begin{enumerate}
\item {\it From the difference in effective lifetime between decays to
    $CP$-eigenstates and decays to flavour-specific (or
    quasi-flavour-specific) final states.}
\end{enumerate}
This is the approach used by BaBar~\cite{Aubert:2003hd,Aubert:2004xga}, where
the $CP$-eigenstate sample is dominated by $J/\psi\,K_S^0$, 
and the quasi-flavour-specific sample is dominated by $D^{(*)-}h^+$ 
($h = \pi,\rho,a_1$).
Semileptonic decays ($D^{(*)-}l^+\nu$) have also been proposed as a
high-statistics flavour-specific control sample~\cite{Dighe:2001gc}.
Flavour-specific final states have ${\cal A}^{\Delta\Gamma}_f = 0$ (the
equality is only approximate for quasi-flavour-specific states),
while ${\cal A}^{\Delta\Gamma}_{J/\psi\,K_S^0} = \cos 2\beta$ to a good
approximation in the SM.

However, if the reconstruction and vertexing requirements differ between the
two samples, there is potential for systematic biases.
An ideal approach is therefore to compare the lifetime distributions of
$J/\psi\,K_S^0$ and the topologically similar but flavour-specific final state
$J/\psi\,K^{*0}$, with $K^{*0} \to K^+\pi^-$.
Since these final states are advantageous for hadron colliders, 
extremely high statistics will be available in the near future from LHCb
(see, for example, Refs.~\cite{Amato:2009zz,Calvi:2009zza}).
Note the strong analogy with the measurement of $y_D$ from the comparison of
effective lifetimes in $D^0\to K^+K^-$ and 
$D^0 \to K^-\pi^+$~\cite{Staric:2007dt,Aubert:2007en,:2009ck}.
\begin{enumerate}
  \setcounter{enumi}{1}
\item {\it From the difference in effective lifetime between decays to
    suppressed and favoured final states.}
\end{enumerate}
Another method which has been used to powerful effect to determine charm
mixing parameters is the comparison of suppressed ($D^0 \to K^+\pi^-$) to
favoured ($D^0 \to K^-\pi^+$) 
decays~\cite{Zhang:2006dp,Aubert:2007wf,:2007uc}.  
A tagged analysis is needed to disentangle the contributions and maximise the
statistical sensitivity.
The analogy in the $B_d^0$ system is the $D^{(*)\pm}\pi^\mp$ final states.
However, while the charm system provides efficient and effective tagging using
the decay $D^{*+} \to D^0 \pi^+$, there is no equivalent for the $B$ mesons.
Therefore, this approach does not appear statistically competitive.
It is, however, worth noting that terms involving $y_d$ should be considered
in order not to bias measurements of the combination of CKM Unitarity Triangle
angles $\sin(2\beta+\gamma)$ from these
decays~\cite{Aubert:2005yf,Aubert:2006tw,:2008kr,Ronga:2006hv}.
\begin{enumerate}
  \setcounter{enumi}{2}
\item {\it From the difference in effective lifetime between $CP$-even and
    $CP$-odd components of self-conjugate vector-vector final states.}
\end{enumerate}
Final states that differ only by having opposite $CP$-eigenvalues have
opposite values of ${\cal A}^{\Delta\Gamma}_f$.  
Hence one possibility is to compare the lifetime distributions between, say,
the $J/\psi\,K_S^0$ and $J/\psi\,K_L^0$ final states.
An alternative is to use angular analysis to disentangle the components with
each $CP$-eigenvalue in self-conjugate vector-vector final states.
This is the approach which is being used to measure $\Delta \Gamma_s$ in
$B_s^0 \to J/\psi \, \phi$ decays~\cite{Aaltonen:2007gf,:2008fj}.
The equivalent final state for the $B_d^0$ is $J/\psi \, K^{*0}$, where
the subsequent decay $K^{*0} \to K_S^0 \pi^0$ must be used.
The requirement on the $K^{*0}$ decay unfortunately complicates the analysis
due to the reduced statistics and the less clean final state
(note, however, that this decay chain has been used for measurements of
$\cos(2\beta)$~\cite{Aubert:2004cp,Itoh:2005ks}).
Analyses using $B^0_d \to D^{*+}D^{*-}$ or $J/\psi\,\rho^0$ are possible but
are not particularly attractive from the point of view of statistics.
\begin{enumerate}
  \setcounter{enumi}{3}
\item {\it From the difference in effective lifetime between $CP$-even and
    $CP$-odd components of self-conjugate multibody final states.}
\end{enumerate}
The contributions from amplitudes with each possible $CP$-eigenvalue for
decays to self-conjugate multibody final states can be disentangled using
Dalitz plot analysis.  However, a tagged analysis is necessary to maximise the
statistical sensitivity.
In the charm system, such analyses have been carried out using
$D^0\to K_S^0\pi^+\pi^-$~\cite{Asner:2005sz,Abe:2007rd,delAmoSanchez:2010xz}. 
Appropriate channels that correspond to $b \to c \bar{c} s$ transitions
include $D^+D^-K^0_S$ and $D^0\bar{D}^0K_S^0$, but the high multiplicity of
the final states reduces the available statistics~\cite{Aubert:2003jq}.
Larger event yields are or will be available in $B_d^0$ decay channels
such as $K_S^0K^+K^-$~\cite{:2008gv,Nakahama:2010nj},
$K_S^0\pi^+\pi^-$~\cite{:2008wwa,Aubert:2009me} and $D_{CP}\pi^+\pi^-$ (where
$D_{CP}$ denotes that the neutral $D$ meson must be reconstructed in a $CP$
eigenstate).  However, the lower branching fractions of these decays compared
to the $b \to c \bar{c} s$ transitions, as well as the inevitable systematic
uncertainties due to Dalitz plot model dependence, make these analyses less
attractive for the measurements of $y_d$.
Note, that these final states are, however, interesting for determinations of
$\cos(2\beta)$~\cite{Charles:1998vf,Latham:2008zs}.

The most promising approach to determine $\Delta \Gamma_d$ therefore appears
to be from the difference in effective lifetimes between decays to the
$CP$-eigenstate $J/\psi\,K_S^0$ and the flavour-specific final state
$J/\psi\,K^{*0}$.
If the vertex position is reconstructed identically (specifically, using only
the $J/\psi$ decay products, typically $\mu^+\mu^-$) in the two cases then the
largest potential source of systematics should cancel almost exactly.
Care will be required to ensure that reconstruction and selection requirements
on the $K_S^0\to \pi^+\pi^-$ and $K^{*0}\to K^+\pi^-$ do not induce a bias on
the lifetime distributions, but this does not appear to present a significant
obstacle to the analysis.
Similarly, potential systematic effects due to the different background
composition in the two channels should be manageable.
One potentially dangerous systematic effect would be an asymmetry, 
$a_{\rm prod}$, between the production rates of $\bar{B}^0_d$ and $B^0_d$.
If non-zero, the cancellation of terms in Eqs.~(\ref{B0barRate})
and~(\ref{B0Rate}) will not be exact, so that the untagged rate of
Eq.~(\ref{untaggedRate}) will include an additional factor in the parentheses
of
$a_{\rm prod} \left( 
  S_f \sin(\Delta m_d \Delta t) - C_f \cos(\Delta m_d \Delta t) 
\right)$.
This can be handled either by fixing $a_{\rm prod}$ based on independent
control samples, or by allowing the presence of such a term in the fit to data.
Asymmetries in the reconstruction of $J/\psi\,K^{*0}$ events (or direct $CP$
violation in the decay to this final state) would similarly introduce a
$\cos(\Delta m_d \Delta t)$ dependence in the lifetime distribution, which can
be handled in the same way.

Finally, the approximation of neglecting corrections arising from $CP$
violation in mixing in Eq.~\ref{B0barRate} and Eq.~\ref{B0Rate} should be
reconsidered. 
Although this is justified for current analyses (since existing experimental
measurements on $CP$ violation in $B^0_d$--$\bar{B}^0_d$
mixing~\cite{Aubert:2006nf,Nakano:2005jb} place much stronger constraints than
those on $\Delta \Gamma_d$~\cite{Aubert:2003hd,Aubert:2004xga}), future precise
measurements will need to take both effects into account.  
Indeed, since the motivation for this study is the possible existence of
anomalous $CP$ violation effects, it would clearly be preferable to perform
an analysis which allows such terms to be non-zero.
The relevant fomulae can be found in
Refs.~\cite{Dunietz:2000cr,Anikeev:2001rk} -- the effect on the untagged $CP$
eigenstate time-dependent decay rate is similar to that of a production
asymmetry indicating the need to determine these parameters from independent
measurements.

In summary, the parameter $\Delta \Gamma_d$, which describes the width
difference between the eigenstates of the effective weak Hamiltonian in the
$B_d^0$--$\bar{B}_d^0$ system, is of interest to test the Standard Model and
potentially to corroborate the recent evidence for anomalous effects in the
$B$ system.
Few measurements exist, and current experiments have the potential to improve
the existing bounds.
Further improvement will be possible at the LHCb experiment, where a newly
proposed method based on the difference between lifetime distributions for the
untagged decays $B^0_d \to J/\psi\,K_S^0$ and 
$B^0_d \to J/\psi\,K^{*0}, K^{*0}\to K^+\pi^-$ appears to be the most
promising approach to reach high precision.

I am grateful to Sheldon Stone and Jacques Lefrancois for comments that
provoked this work, 
and to Paul Harrison, Ulrich Nierste and Guy Wilkinson 
for comments on the manuscript.
Thanks also to Yasmine Amhis and Greig Cowan for pointing out a typo in
Eqs.~(\ref{sinhApprox}) and~(\ref{untaggedRate}).
This work is supported by the 
Science and Technology Facilities Council (United Kingdom) and by the 
European Research Council under FP7.

\end{document}